\documentclass[conference]{IEEEtran}

\makeatletter

\def\ps@IEEEtitlepagestyle{%
  \def\@oddfoot{\mycopyrightnotice}%
  \def\@evenfoot{}%
}
\def\mycopyrightnotice{%
  {\footnotesize 978-1-6654-7087-2/22/\$31.00~\copyright~2022 IEEE\hfill}
  \gdef\mycopyrightnotice{}
}

\usepackage{url}
\usepackage{blindtext}
\usepackage{eso-pic}
\IEEEoverridecommandlockouts
\usepackage{cite}
\usepackage{amsmath,amssymb,amsfonts}
\usepackage{algorithmic}
\usepackage{graphicx}
\usepackage{textcomp}
\usepackage{xcolor}
\def\BibTeX{{\rm B\kern-.05em{\sc i\kern-.025em b}\kern-.08em
    T\kern-.1667em\lower.7ex\hbox{E}\kern-.125emX}}
    
\usepackage{eso-pic}
\newcommand\AtPageUpperMyright[1]{\AtPageUpperLeft{%
 \put(\LenToUnit{0.17\paperwidth},\LenToUnit{-2cm}){%
     \parbox{0.9\textwidth}{\raggedleft\fontsize{8}{11}\selectfont #1}}%
 }}%
\newcommand{\conf}[1]{%
\AddToShipoutPictureBG*{%
\AtPageUpperMyright{#1}
}
}   
\begin{document}
\title{\vspace*{1cm} Combating harmful Internet use with peer assessment and differential evolution\\
}

\author{\IEEEauthorblockN{1\textsuperscript{st} W.W. Koczkodaj}
\IEEEauthorblockA{\textit{Department of Computer Science} \\
\textit{Laurentian University}\\
Sudbury, Canada \\
wkoczkodaj@cs.laurentian.ca}
\and
\IEEEauthorblockN{2\textsuperscript{nd} M. Mazurek}
\IEEEauthorblockA{\textit{Department of Complex Systems} \\
\textit{Rzeszów University of Technology}\\
Rzeszów, Poland \\
miroslaw.mazurek.rzeszow@gmail.com}
\and
\IEEEauthorblockN{3\textsuperscript{rd} W. Pedrycz}
\IEEEauthorblockA{\textit{Department of Electrical and Computer Engineering} \\
\textit{University of Alberta}\\
Edmonton, Canada \\
wpedrycz@ualberta.ca}
\and
\IEEEauthorblockN{4\textsuperscript{th} E. Rogalska}
\IEEEauthorblockA{\textit{Institute of Pedagogy} \\
\textit{University of Szczecin}\\
Szczecin, Poland \\
elzbietarogalska@o2.pl}
\and
\IEEEauthorblockN{5\textsuperscript{th} R. Roth}
\IEEEauthorblockA{\textit{Laurentian Univesity} \\
Sudbury, Canada \\
reuben.roth@gmail.com}
\and
\IEEEauthorblockN{6\textsuperscript{th} D. Strzalka}
\IEEEauthorblockA{\textit{Department of Complex Systems} \\
\textit{Rzeszów University of Technology}\\
Rzeszów, Poland \\
strzalka@prz.edu.pl}
\and
\IEEEauthorblockN{7\textsuperscript{th} A. Szymanska}
\IEEEauthorblockA{\textit{Institute of Psychology} \\
\textit{Cardinal Stefan Wyszynski University}\\
Warsaw, Poland \\
elysium567@gmail.com}
\and
\IEEEauthorblockN{8\textsuperscript{th} A. Wolny-Dominiak}
\IEEEauthorblockA{\textit{University of Economics in Katowice} \\
Katowice, Poland \\
alicja.wolny-dominiak@ue.katowice.pl}
\and
\IEEEauthorblockN{9\textsuperscript{th} M. Woodbury-Smith}
\IEEEauthorblockA{\textit{Biosciences Institute} \\
\textit{Newcastle University}\\
Newcastle, United Kingdom\\
woodbur@mcmaster.ca}
\and
\IEEEauthorblockN{10\textsuperscript{th} O.S. Xue}
\IEEEauthorblockA{\textit{Department of Computer Science} \\
\textit{Laurentian University}\\
Sudbury, Canada\\
sxue@laurentian.ca}
\and
\IEEEauthorblockN{11\textsuperscript{th} R. Zbyrowski}
\IEEEauthorblockA{\textit{Department of Operational Research} \\
\textit{University of Warsaw}\\
Warsaw, Poland\\
zbyrowskirafal@gmail.com}
}

\maketitle
\conf{\textit{  Proc. of the International Conference on Electrical, Computer and Energy Technologies (ICECET 2022) \\ 
20-22 June 2022, Prague-Czech Republic}}

\begin{abstract}
Harmful Internet use (HIU) is a term coined for
	the unintended use of the Internet. In this study, we propose
	a more accurate HIU measuring method based on the peer
	assessment and differential evolution approach. The sample data
	comprises a juvenile population in Poland; 267 subjects 
	assessed 1,513 peers. In addition to classic statistical analysis,
	differential evolution has been employed. Results indicate that
	there may be a substantially higher rate of HIU than other studies
	have indicated. More accurate measurement of the adolescent
	population influx affected by HIU is needed for healthcare
	and welfare system planning.
\end{abstract}


\section{Introduction}
When used inappropriately, the harmful Internet use (HIU) becomes a major problem, especially for the younger generation.
According to \cite{IHGetal2019}, problematic Internet use has been linked to behavioral addiction, major depressive disorder, ADHD, deficit/hyperactivity disorder, sleeping disorders, cognitive deficits, and suicides.
Terms like "Internet addict" have commonly been used to recognize the burgeoning destructive potential of the excessive Internet use or being attracted to illicit pastimes. 

In this study, a rating scale is utilized as a tool for quantitative measuring of a new social phenomenon: HIU. Considerable efforts are being made to improve its accuracy and reduce the uncertainty of measurements when enhanced by the peer assessment and analyzed with the help of differential evolution.

The Netflix September 2020 release of the docudrama “The Social Dilemma” (see \cite{SocialDilemma}) has generated considerable publicity related to HIU, and is a prime illustration of the need for improved measurement of HIU.

The Social Dilemma is a pivotal docudrama (see \cite{SocialDilemma}) which delves into the dangers of social media in particular. By interviewing the designers of social media platforms, the film makes a compelling case that social media poses a viable threat to civilization itself. Social media companies constrain us into adopting modes of thinking and behaving in ways that are profitable for corporations, rather than thinking and behaving in ways that are based on our own goals, beliefs, or values. Designers of social media platforms force us to give our time away to corporations selling 'big data' to their clients. We argue that the approach presented here should be used to measure the harm suggested in The Social Dilemma. Group assessments can be improved by: 
\cite{JK1996}, \cite{BFK2011}, \cite{KO1997}, and \cite{FKS2010}. Another type of harmful Internet use is presented by \cite{SIR}.

\section{Methods}
Our intensive search of the most recognized subscription-based scientific citation indexing services: Scopus (by Elsevier, Netherlands) and Web of Knowledge (by Clarivate Analytics, USA) has traced the first formal introduction of the scientific peer review method to the 17th century. Henry Oldenburg (also Henry Oldenbourg) FRS, was a German theologian known as the proponent of scientific peer review.

The American Medical Association uses the medical peer review in its assessment of the improving quality and safety process in health care organizations as well as in the process of assessing the clinical behavior and compliance with professional society membership standards (see \cite{AMA}).

The self-assessment study (see\cite{PBS2017}) was a survey of more than 250,000 individuals 45 years of age or older residing in New South Wales (NSW), the most populous state of Australia. It has demonstrated that individuals are significantly (by 36.5\%) more likely to under-report mental disorders. Evidently, the self-assessment should not be regarded as a reliable measurement method. 

Compared with mobile device recordings, most parents underestimated (35.7\%) or overestimated (34.8\%) their child's use. The total inaccuracy (70.5\%) supports the common knowledge that the vast majority of parents miscalculate their children's time spent using mobile devices.

\subsection{Data acquisition and questionnaires}

In this study, 267 participants took part. Participants evaluated 1,518 of their peers. 185 of 267 subjects were elementary school students. They evaluated 1,238  peers. In addition to this, 82 parents evaluated 280 children, their own children, or children they knew. The age of the assessed participants was between 10 and 22 years. 
Secondary school students, aged from 12 to 15, have provided answers to questions in two questionnaires. The children's parents were encouraged to take part in this study.

One questionnaire required the respondents to provide a list of close friends and family members close to their age. Parents, grandparents, girlfriends, boyfriends, and/or guardians as they are other closely related (by an emotional connection with the subjects for an unbiased assessment) were excluded from this list. Subjects were instructed to list everyone who they could evaluate (not just those who you may suspect of HIU). Friendship or any kind of professional (e.g., study) relationship was regarded as acceptable but more intimate relationships were not since it may impair the objectivity of assessment.
The second questionnaire was used to measure the HIU of the peers listed in the first questionnaire. It is discussed in Subsection~\ref{Figure67}.

A small scale rating, investigated in \cite{FKS2010}, has been used. In the case of children, the scale 0 to 3 is easier to comprehend and use. The meaning of 0 usually signifies absence or lack of knowledge and 3 stands for the maximum of some quality or knowledge about the subject. Modeling was carried out separately for answers obtained from the group of children, university students and parents.

\section{Differential evolution meta-heuristic}
Differential evolution (DE) was introduced in \cite{SRDE}. It has been used in \cite{KW2019} for finding a solution by iterative improvement of a candidate solution (e.g., weights) for a given objective function $f$ which was are under the curve (AUC) of the receiver characteristic curve (ROC). 

DE is widely recognized as one of the most powerful meta-heuristics (sometimes called an algorithm) that is based on the classic developmental process used for evolutionary algorithms (EAs). When compared with other  traditional approaches, such as EAs, DE uses the scaled differences of vectors to produce new candidate solutions in the population. Hence, no separate probability distribution needs to be used to perturb the population members \cite{SRDE}. DE is also characterized by the advantages of having few parameters leading to the simplicity of implementation. 

\noindent DE could be specified by the following steps:
\begin{enumerate}
	\item \textit{initialization} - An initial population that is sampled uniformly at random within the search bounds is created.
	\item \textit{mutation} - Three components namely mutation, crossover and selection are adopted to evolve the initial population.
	\item \textit{crossover} - The mutation and crossover are used to create new solutions.
	\item \textit{selection} - The selection determines the solutions that will breed a new generation is made.
\end{enumerate}

DE remains inside a loop until the stopping criterion is met. Each step is explained separately in subsequent subsections.
From the syntax point of view, it looks like an algorithm but it is a heuristic since proving its convergence is a challenge for a general case.

\subsection{Initialization}\label{AA}
Like other optimization algorithms, DE starts with a randomly initialized population of order NP. Parameter vectors are called individuals. Each such individual represents a $D$-dimensional vector of decision variables (parameters). The {\it i-th} individual for the generation $G$ is denoted as follows:
\begin{equation}
\mathbf{x_i}^G=[x_{i,1}^G, \ldots, x_{i,D}^G],
\end{equation}
where $j=1\ldots,D$ and $i=1,\ldots,\mathrm{NP}.$\\
Both upper and lower bounds of the decision variables should be restricted to their minimum and maximum for $i = 1,\ldots,NP$ before the population is initialized.
\begin{eqnarray*}
	\mathbf{B_L}=[B_{L,1}, \ldots ,B_{L,j}]=[\min x_{i,1}^G, \ldots ,\min x_{i,D}^G],\\
	\mathbf{B_U}=[B_{U,1}, \ldots ,B_{U,j}]=[\max x_{i,1}^G, \ldots,\max x_{i,D}^G]
\end{eqnarray*}
Once initialization search ranges have been determined, DE assigns each decision variable $j$ of $ith$ individual a value from within the specified range as follows \cite{SRDE}, \ for $G=0,$
\begin{equation}\
x_{i,j}^0 = B_{L,j}+r(B_{U,j}-B_{L,j}),
\end{equation}
where $r\in[0,1]$  represents a uniformly distributed random number within the range $0\leq r <1$.

\subsection{Mutation}
After initialization, the mutation operator produces new solutions by forming a mutant vector (trial vector) for each parent individual (target vector). For each target vector, its corresponding trial vector can be generated by different mutation strategies. Each strategy employs different approaches to make a balance between the exploration and exploitation tendencies. For {\it the i-th} target vector at the {\it G} generation the five most well-known mutation strategies are presented as follows: 
$r1, r2, r3, r4, r5\in\mathrm{NP}$ are five different randomly generated integer numbers. Furthermore, $F$ is a scaling factor $\in[0,2]$ affecting the difference vector and $\mathrm{best}\in\mathrm{NP}$ is an index of the best individual vector at generation $G.$
\begin{itemize}
	\item[(1)] DE/rand/1
	\begin{equation*}
	\mathbf{v_i}^G = \mathbf{x_{r1}}^G+F\cdot(\mathbf{x_{r2}}^G-\mathbf{x_{r3}}^G) ,
	\end{equation*}
	\item [(2)] DE/best/1
	\begin{equation*}
	\mathbf{v_i}^G = \mathbf{x_{best}}^G+F\cdot(\mathbf{x_{r1}}^G-\mathbf{x_{r2}}^G) ,
	\end{equation*}
	\item [(3)] DE/rand-to-best/1
	\begin{equation*}
	\mathbf{v_i}^G = \mathbf{x_{i}}^G+F\cdot(\mathbf{x_{best}}^G-\mathbf{x_{i}}^G)+F\cdot(\mathbf{x_{r1}}^G-\mathbf{x_{r2}}^G) ,
	\end{equation*}
	\item[(4)] DE/best/2
	\begin{equation*}
	\mathbf{v_i}^G = \mathbf{x_{best}}^G+F\cdot(\mathbf{x_{r1}}^G-\mathbf{x_{r2}}^G)+F\cdot(\mathbf{x_{r3}}^G-\mathbf{x_{r4}}^G) ,
	\end{equation*}
	\item[(5)] DE/rand/2
	\begin{equation*}
	\mathbf{v_i}^G = \mathbf{x_{r1}}^G+F\cdot(\mathbf{x_{r2}}^G-\mathbf{x_{r3}}^G)+F\cdot(\mathbf{x_{r4}}^G-\mathbf{x_{r5}}^G) ,
	\end{equation*}
\end{itemize}

\subsection{Crossover}
In this step, DE applies a discrete crossover approach to each vector and a mutant vector. The basic version of DE incorporates the binomial crossover defined as follows \cite{SRDE}:
\begin{equation*}
\mathbf{u_i}^G =
\begin{cases}
{v}_{i,j}^G \text{ if }(r \leq CR)\text{ or } (j=j_{\mathrm{rand}}) \\
{x}_{i,j}^G \text{ otherwise,}
\end{cases}
\end{equation*}
\noindent where:\\
$CR$ is the user-specified crossover rate which determines the probability of mixing between vectors and mutant vectors,\\
$j_{rand}\in[0,D]$ is a randomly picked integer number.

\subsection{Selection}
In this step, DE adopts a selection mechanism to choose the best individuals according to their fitness for producing the next generation of population. Toward this goal, it compares performance of the trial and target vectors and copies the better one into next generation; as presented above.
\begin{equation*}
\mathbf{u_i}^{G+1} =
\begin{cases}
\mathbf{u_i}^G \text{ if } f(\mathbf{u_i}^G)\leq f(\mathbf{x_i}^G)
\\ \mathbf{x_i}^G \text{ otherwise,}
\end{cases}
\end{equation*}
where, $f$ is the objective function that should be optimized.
\section{Measurement model}
\label{MM}
The population selection followed two common sense rules, ``as random as possible'' with ``as many subjects as is feasible''. Respondents provided answers about each assessed individual HIU patterns after the concept of HIU was addressed to them in the class and the presence of a teacher.

\begin{enumerate}
	\item Q1. I know his/her HIU pattern (N/A?).
	\item Q2. He/she prefers HIU than socializing. 
	\item Q3. His/her acquaintances and/or parents are concerned about his/her HIU. 
	\item Q4. HIU impair his/her health, hygiene and eating pattern. 
	\item Q5. He/she avoids other activities.
	\item Q6. He/she tried to decrease HIU but failed.
	\item Q7. HIU negatively impacts his/her school performance. 
	\item Q8. Rating of his/her HIU as (0 = never plays, 3 = extremely harmful).
	\item Q9. My sex is: F -- female, M -- male, U - undeclared.
\end{enumerate}

\begin{figure}[]
	\centering
	\includegraphics[scale=0.45]{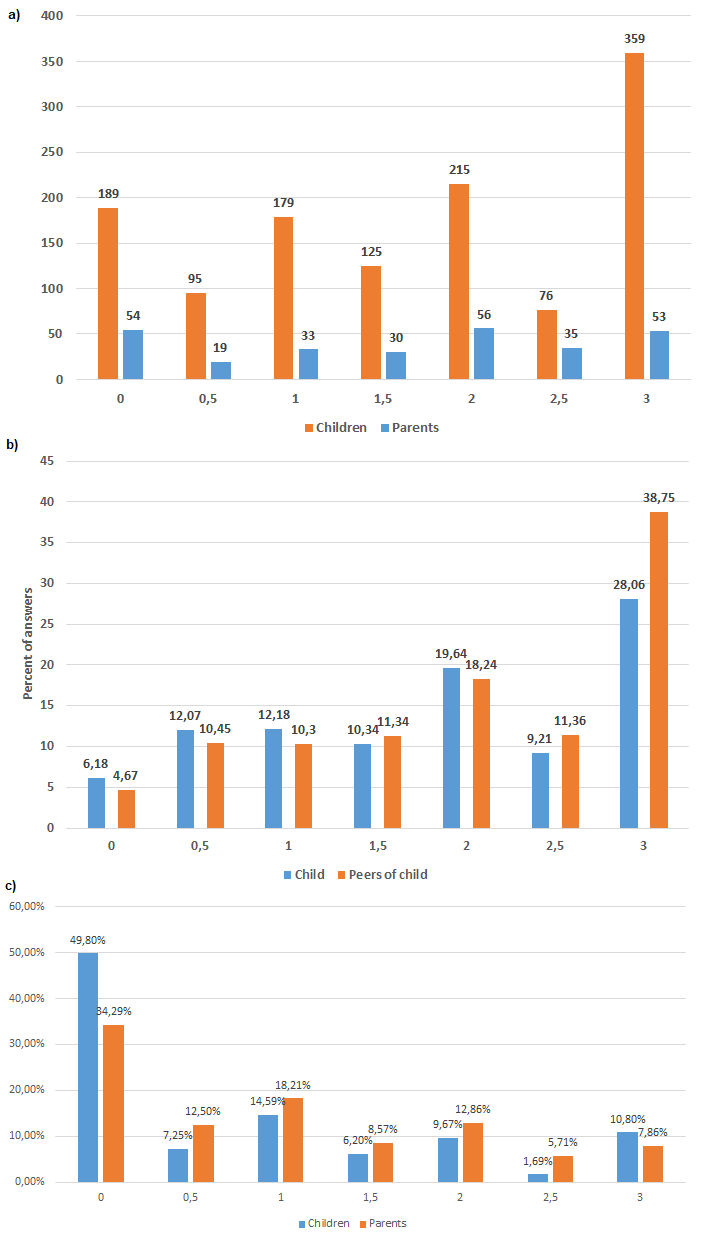}
	\caption{a) Number of answers for question Q1: "I know his/her HIU pattern(N/A?)"; b) Q8 -- Analysis of avoiding other activities by children and parents; c) Q7 -- HIU negatively affects his/her school performance}
	\label{Figure2}
\end{figure}

\section{Results}
Poland is the 6th largest economy in the European Union and 21st in the World. With a Human Development Index of 0.865, Poland is in 33rd place on the list and regarded as one of the ``very high human development'' countries. For these reasons, Poland is representative of economic and social points of view as a developed country. 

In a group of elementary school students, the respondents evaluated an average of 6 peers each. The smallest number of peers that the respondent evaluated was N=1 and the largest was 24. In the parents' group, the average was 3 (range 1 to 10) children evaluated, the smallest number of evaluations per parent is N=1, the largest was 10 evaluations.

A detailed analysis of Q1 shows that 52\% of respondents know the HIU patterns of assessed subjects.
One of the important aspects of human growth is the socialization process. Social development ensures a safe and healthy relationship with individuals. A creative use of the Internet can have the effect of reinforcing a sense of friendship and connections for teens who play online games with friends. The results obtained in the survey (question Q2) do not support it. Above 61\% of children prefer socializing (in analysis the answers with a weight 1.5 -- 3.0 were added). 

Analysis of question Q2 shows that 39\% of children see a problem related to avoiding social contacts. An analysis of children's behavior by their parents shows that almost 52\% of parents know their children's HIU pattern but do not regard it as a potential problem as illustrated by Fig.~\ref{Figure2} "A". Parents admitted their knowledge of the violent content to which their children were exposed. Parents (53\%) notice isolation of their children from peers and society.

These results suggest that the problem of HIU is not well recognized, despite answers to other related questions indicating HIU problems.
The negative opinion threshold value was set to 2, hence the totals were obtained as the sum of answers with 2 and 3.

Q8 was subjected to a more detailed analysis. It turns out that when analyzing HIU, parents assess the situation as much worse in other children (67\%) than at their own (57\%) as shown in Fig.~\ref{Figure2} "b". 

Fig.~\ref{Figure2} "c" shows the analysis of question Q7. Nearly 60\% of children and 50\% of parents do not see a problem related to HIU. Negatively effects of HIU on school performance is seen as a problem by only 26\% of parents and 22\% of children.

\begin{figure}[]
	\centering
	\includegraphics[scale=0.6]{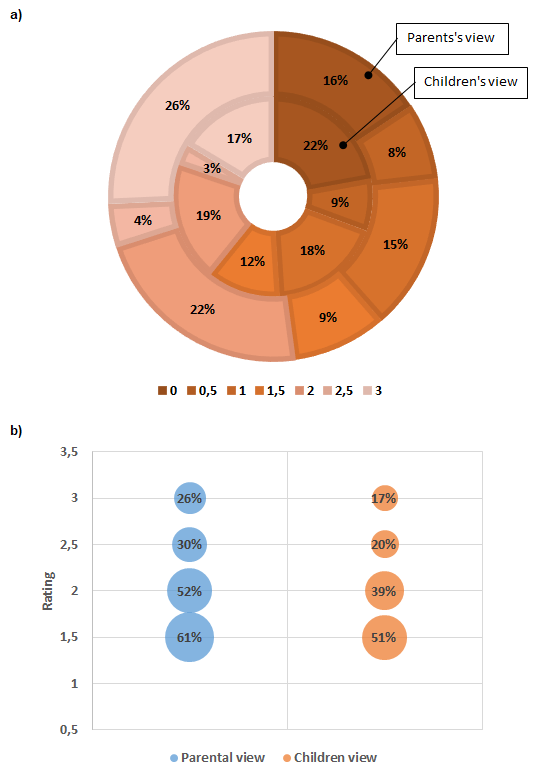}
	\caption{a) Preferences between HIU and socializing -- the answers for question Q8; b) Answers for scales in groups -- question Q8.}
	\label{Figure3}
\end{figure}

An analysis of question Q3 shows that above 65\% of respondents and 38\% of parents consider HIU as a normal activity. They are not concerned about their HIU via social network. 
Only 32\% of parents regarded HIU as a negative activity.

According to the respondents, there is no problem with the  deterioration of health or hygiene due to the HIU. The analysis of question Q4 
shows that almost 78\% of children and 54\% of parents believe that HIU does not impair their health, hygiene and eating pattern. Only 19\% of parents regard HIU as a contributing factor for deterioration of health or hygiene in their children.

Analysis of question Q5 shows that, consistent with the previous question, nearly 67\% of children and 46\% of parents do not view as a problem of the avoidance other activities as negative to the child's development. Abuse of games and the Internet is seen as a problem by only 24\% of parents and is viewed as a lower risk by children (only 15\%).

The results of the analysis of question Q6 show that nearly 73\% of children and 59\% of parents have not even attempted to reduce their HIU. The majority of respondents (56\% of children, 47\% of parents) believe that HIU has no application to them or their children. Only 13.5\% of parents stated that the school performance of their children had deteriorated, while 22\% of children expressed concerns about their peers.

Fig.~\ref{Figure3} "a" shows the percentage of answers to question Q8: "I rate his/her HIU as (0 = never plays, 3 = extremely harmful)". The inner ring reflects the children's opinion. The external ring shows parents' opinions. 

The bubble chart in Fig.~\ref{Figure3} "b" shows the analysis of HIU. The largest bubble shows the population percent for the score of 1.5 (moderate) to 3 (extremely harmful). Parents have evaluated children (their own and peers of their children). 61\% of parents regard HIU as concerning. Elementary school children evaluated other peers' HIU at 51\%. The next level of the bubble chart is for the accumulated total from 2 to 3. It is 52\% for the parental view and 39\% for elementary school children. For the accumulated total 2.5 and 3, the percentage is 30\% and 20\% respectively. For the extreme case of HIU (3), it is 26\% and 17\% respectively.
\begin{figure}[]
	\centering
	\includegraphics[scale=0.3]{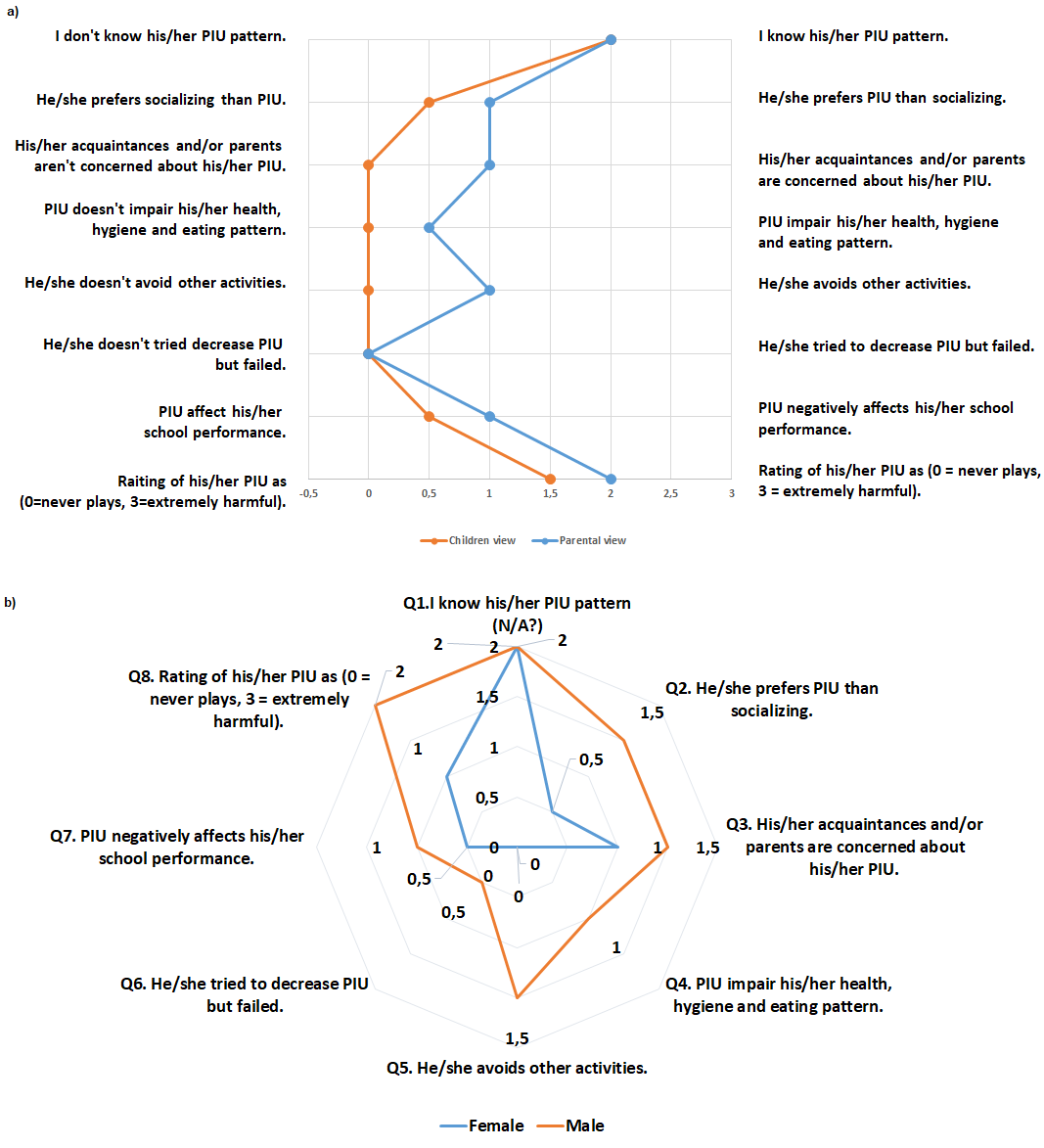}
	\caption{a) Semantic differential for HIU parents and children opinions (medians); b) HIU Male/Female: parental opinions (median).}
	\label{Figure67}
\end{figure}
Fig.~\ref{Figure67} "a" shows the dispersion between responses of parents and children. Fig. \ref{Figure67} "a" indicates that responses of parents are more negative especially for questions 3 and 5. 
Fig.~\ref{Figure67} "b" shows dispersion of ratings provided by parents for the female and male population. These ratings are for question 1. They demonstrate that in the opinion of parents, both sexes are above an assumed neutral point of 1.5. However, question 8 was evaluated by parents in a way indicating that male children have more problems than female children. The difference in responses to question 5 is interesting. It indicates that the male population of children avoids other activities more often than the female population. There are three levels of difference between sexes which are regarded as substantial.

\subsection{Statistical model specification}
In the proposed statistical model, a dichotomous dependent variable $Y$ takes only two values: 0 and 1. It is a binary (or dichotomous) choice model. The relationship between $y$ and the explanatory variable consists of modeling the probability for the i-th object of: 

\[
p_i=F(x_i^{'}\beta{})
\]

\[
x_i^{'}\beta{}={(1\ X_{1i}\ X_{2i}... \ X_{ki})({\beta{}}_0\ {\beta{}}_1\
	{\beta{}}_2...\ {\beta{}}_k)}^{'}
\]

$X_{ki}$ -- variable explaining the number $k$ for the observation $i$;

$\beta_k$ --  parameter for the variable explaining the number $k$.\\

The object of our modeling is a hidden variable $y*$, whose values are not observed. The hidden variable $y*$ represents a tendency of an observation unit to make a decision or to take a state corresponding to $y = 1$. 
It is assumed that if this tendency is positive, then $y = 1$. Thus, $y = 1$ if $y* > 0$, and $y = 0$ if $y* \le 0$. The inclination $y*$ is the following function of the model explanatory variables:

\[
y_i^*=x_i^{'}\beta{}+e_i
\]

$e_i$ -- the error of the model (\textit{white noise}). \\

The ordered logit models correspond to the ranking scale of the dependent variable in questions Q6 and Q7. In such models, the explained variable takes discrete values, ordered in a natural way (e.g. 0, 1, 2, 3, 4, 5, 6). Formally, it is assumed that the ordered variable $y$ is a limited record of a continuous variable $y*$. 

In this study, six models have been defined: 
\begin{itemize}
	\item M1. Estimations for a group of children (Q6).
	\item M2. Assessment of the group of children (Q7). 
	\item M3. The high HIU rating implies a male (Q9).
	\item M4. Estimations for a group of parents (Q6). 
	\item M5. Parents' opinion (Q7).
	\item M6. HIU in the group of children (Q9).
\end{itemize}

It is worth noting that the number of correctly predicted cases exceeds 50\% for almost all estimated models. The exception is model M5 with only 42.6\% correctly predicted cases. In addition, the ''likelihood ratio test'' for each model allows us to reject the null hypothesis of the equality to 0 for all coefficients. In model M1, all statistically significant exogenous variables have a positive impact on the assessment obtained in question Q6. This means that the high rating obtained in any of the exogenous questions in model M1 (i.e., Q3, Q4, Q5 or Q7) increases the chance that the person has unsuccessfully tried to quit an unwanted activity. In this model, the care rating of parents or acquaintances (Q3) has the strongest positive impact on the assessment obtained in question (Q6).

In model M2, the positive relation of question Q7 with the majority of regressors is also evidenced. The only exception is the gender (Q9). Gender reduces the chance of school-related problems connected to HIU (Q7) for females. The relationship between gender and the HIU rating obtained in the question (Q8) is shown by the logit binary model.

For parents' (model M4), (Q6) infers that there is the largest positive dependence between ''the stop HIU question'' (Q6) and the avoidance of other activities (Q5). Therefore, avoiding other activities (Q5) can be an important predictor of excessive gaming. The parents' increased concern (Q3) and the fact that the respondent knows that the person has a HIU problem (Q1) can be an important predictor of HIU. In the parental perception (model M5), females suffer much less from school problems as a result of HIU (Q9). However, the high rating obtained in the HIU (Q8) strongly implies educational problems (Q7) for both genders. A high impact on school problems also affects avoiding other activities (Q5). A slightly smaller impact on rating (Q7) has a reduced hygiene and food eating pattern (Q4) and poorer interpersonal relationship (Q2).

Model M6 shows the relationship between the intensity of the HIU (Q8) and the subject's gender (Q9). The analysis has been carried out for the answers obtained by parents. The conclusion is similar to the answers obtained for the group of children. The high score recorded in the area of HIU (Q8) significantly reduces the chance that the examined person is female.

\section{Conclusions}

By using the proposed measurement enhancement, our study has indicated that the HIU penetration is at a much higher level among children in Poland than we have previously realized.  

Our findings are consistent with common sense and observations that using measurements based on assessments by peers we gain in accuracy when compared to self-assessments or parental assessments. Further improvement of accuracy is expected to be gained by adding approaches in \cite{K1996, K1998} in the followup publication.

The presented models show a strong correlation between
HIU and avoiding other activities, such as sports and live
socializing. Poorer levels of hygiene, health, and nutrition
pattern can also be, in part, attributed to HIU based on the
results of our study. Additionally, gaming often raises the
anxiety of acquaintances and/or parents. According to \cite{ER2021}: ``How does the school limit children through infrastructure and is it really a limiting phenomenon, or maybe a place of silence will prove to be a good space element for development, for example, spiritual.''

Research in drug abuse and addiction also teaches us that
parents are often the last to know about their children’s
addiction problems. This discouraging situation is exacerbated
by the unreliability of the current measurements of Harmful
Internet use. The proposed peer review approach is a more
objective way of measurement, which seems to be worth
additional research effort as an innovative approach.
Poland has the sixth-largest economy in the European Union
by nominal GDP and the fifth-largest by purchasing power
parity GDP. Poland has been classified as a high-income
economy by the World Bank, ranking 22nd by GDP (nominal)
and 19th worldwide by GDP (PPP). The 2017 Economic
Complexity rank 21 of Poland in the world reflects diverse
strong economy. Poland is a conservative society and it is
reasonable to assume that our results (hence their importance)
are representative for all developed countries.

\section*{Acknowledgment}

The authors would like to thank the Board of Education in Rzesz\'ow (Poland) for allowing us to collect data and for collecting consent forms signed by parents.

\end{document}